# Quasi-one-dimensional metallic conduction channels in exotic ferroelectric topological defects

Wenda Yang,[†a] Guo Tian,[†a] Yang Zhang,[‡§] Fei Xue,[‡] Dongfeng Zheng,[†] Luyong Zhang, Yadong Wang,[†] Chao Chen,[†] Zhen Fan,[†] Zhipeng Hou,[†] Deyang Chen,[†] Jinwei Gao,[†] Min Zeng,[†] Minghui Qin,[†] Long-Qing Chen,[‡] Xingsen Gao,[*,†] and Jun-Ming Liu[†§]

[†]Guangdong Provincial Key Laboratory of Quantum Engineering and Quantum Materials and Institute for Advanced Materials, South China Academy of Advanced Optoelectronics, South China Normal University, Guangzhou 510006, China

[‡]Department of Materials Science and Engineering, The Pennsylvania State University, University Park, Pennsylvania 16802, United States of America

[§]Laboratory of Solid-State Microstructures and Innovation Center of Advanced Microstructures, Nanjing University, Nanjing 210093, China

[*]Corresponding author: xingsengao@scnu.edu.cn

[a]These authors contributed equally

**[Abstract]** Ferroelectric topological objects (e.g. vortices, skyrmions) provide a fertile ground for exploring emerging physical properties that could potentially be utilized in future configurable nanoelectronic devices. Here, we demonstrate quasi-one-dimensional metallic high conduction channels along two types of exotic topological defects, i.e. the topological cores of (i) a quadrant vortex domain structure and (ii) a center domain (monopole-like) structure confined in high quality $BiFeO_3$ nanoisland array, abbreviated as the vortex core and the center core. We unveil via phase-field simulations that the superfine (< 3 nm) metallic conduction channels along center cores arise from the screening charge carriers confined at the core whereas the high conductance of vortex cores results from a field-induced twisted state. These conducting channels can be repeatedly and reversibly created and deleted by manipulating the two topological states via an electric field, leading to an apparent electroresistance effect with an on/off ratio higher than $10^3$. These results open up the possibility of utilizing these functional one-dimensional topological objects in high-density nanoelectronic devices such as ultrahigh density nonvolatile memory.

**Introduction**

Topological objects and defects (e.g. domain walls, vortices, skyrmions) in condensed matter have garnered massive attention as an arena of exploring emerging exotic phenomena and new functionalities [1, 2]. In materials with ferroic order, these topological objects can also be manipulated and controlled by external fields without disrupting their host lattice, making them promising elemental building blocks for potential configurable topological nanoelectronics [3-6]. To this stage, most earlier investigations have hitherto focused on the properties of two-dimensional (2D) defects (namely domain walls). For instance, a wealth of unique properties have been observed in ferroelectric domain walls, *i.e.* two-dimensional (2D) topological defects, including enhanced domain wall conductivities [7-11], enhanced photovoltaics [12], giant magnetoresistances [13], extraordinary magnetisms [14], and quantum oscillation behaviors [15], among many others. These functionalities could underpin a wide range of potential configurable nanoelectronic, magnetoelectronic, and optoelectronic applications, in particular the energy-efficient current-readout nonvolatile memories based on reversibly creating/eliminating the conductive domain walls [16, 17]. However, such conceptual devices usually face the obstacles of low scalability and unstable restoration process due to the difficulty in deterministic control of the domain walls. Recent studies did show the enhanced repeatability of domain wall switching in center-type domains due to the topological and geometric restriction effect [18]. This motivation, in practice, could suffer from the non-uniformity in domain wall conductivity arisen from local distortions (*e.g.* bending or tilting) [19], detrimental to device performance.

In recent years, there has been an increasing interest in more complex ferroelectric topological objects, leading to the discovery of a series of topological states, such as closure domain state [20-24], quadrant vortex states [25-28], circular vortices lattices [29], skyrmions [30], bubble domains [31], and center domain states (monopole-like structure with polarization pointing inward/outward the core) [18,32,33] in size-confined thin films / superlattices / nanostructures, as well as vortex network in improper ferroelectric single crystals (e.g. YMnO$_3$) [14, 34]. These tantalizing findings have kindled the excitement for exploring new device concepts associated with these exotic topological defects. For example, exotic one-dimensional (1D) topological defects, e.g. vortex domain cores or center domain cores, take the advantages of 1D superfine dimensionality and topological protection nature (resilience against perturbations), which not only allows scaling-down the device dimension to nanometer-scale but also substantially improves the restoration repeatability and stability, promising for high-density integrated devices. For instance, ultra-small bi-stable vortices as tiny as 3 nm in diameter, can be stabilized in size-confined nanostructures [29, 35], offering a possibility of developing ultra-dense memory with an areal density of 60 Tb/in$^2$. Currently, interest in 1D topological defects is seeing a rapid development and highly appreciated, nonetheless exotic functionalities of these 1D defects yet remain elusive.

In this work, we demonstrate metallic conduction superfine (< 3 nm) channels in two types of exotic topological defects, namely a quadrant vortex core or simply vortex core and a quadrant center domain core or simply center core, in an array of BiFeO$_3$ (BFO) nanoislands (see Fig. 1a for the nanoislands, Fig. 1c for a vortex state, and Fig. 1d for a center state). Interestingly, these intriguing topological states can be controllably created and eliminated, leading to a remarkable electro-resistance effect with an on/off ratio larger than 10$^3$ and good (thermal and fatigue) stability. The observed phenomena might open a route towards none-destructive ultrahigh density memories, based on these superfine topological objects, and offer an excellent paradigm for the concept of topological electronics (or "topotronics").

**Topological domain structures and conduction patterns**

In this work, BFO nanoisland arrays were directly patterned from high quality epitaxial BFO films (~ 35 nm in thickness) *via* the nano-sphere lithography technique using the polystyrene sphere (PS) arrays as templates [11]. The detailed fabrication process can be found in the Method section and Supplementary Materials (see Section A and Supplementary Fig. 1a). The structure of nanoisland arrays was characterized using atomic force microscopy (AFM), scanning electron microscopy, X-ray diffraction (XRD), and XRD reciprocal space map (RSM, see Fig. 1b), confirming that the BFO islands have the highly epitaxial rhombohedral BFO phase (Supplementary Fig. 1b ~ 1c). The ferroelectricity of randomly selected nanoislands was examined using local piezo-response testing, revealing a butterfly-like amplitude-voltage and phase-voltage hysteresis loops (see Supplementary Fig. 1d).

To examine the domain structures, we conducted a vector piezoresponse force microscopy (PFM) measurement by recording the PFM images upon the in-plane rotation of the sample at 0° and 90° angles with respect to the reference direction. Consequently, the 3D domain structures from the PFM data can be reconstructed, as reported earlier [32, 36]. Here, we choose two representative nanoislands for illustration: one possesses a single vortex domain state, and the other a single center domain state. These domain states were created from the initial wedge-like domain structure by applying suitable bias voltages, and the details are illustrated in Supplementary Fig. 3, while an identification of the two types of domain structures is illustrated in Fig. 1c & 1d. One can derive the in-plane polarization vector maps from the lateral PFM phase images (Lat-Phase) measured for the clockwise rotation of the sample at 0° and 90° angles, which allows a determination of the local polarization components along *x-* and *y-axis* (corresponding to [100]- and [010]-axis of BFO crystal lattice), respectively, using the fact that only eight possible [111] polarization directions are permitted by crystal symmetry of BFO. It was found that the first nanoisland contains four quadrant head-to-tail domains (with upward out-of-plane polarization component, as reflected by the uniform dark-contrast in vertical PFM phase image), forming an in-plane flux-closure vortex structure, along with four 71° neutral domain walls (NDWs) meeting at a core (see Fig. 1c). This is a typical characteristic of the vortex domain structure, as displayed in the schematic vortex structure in Fig. 1c. The second one consists of four quadrant domains (with upward out-of- plane polarization component too), with in-plane polarization components pointing inwards to the center core and four head-to-head CDWs meeting at the core, consistent with the feature of so-called center-convergent topological state [25, 32], as shown in the schematic vortex structure in Fig. 1d.

Hereafter, for the convenience of description, the flux-closure vortex state will be abbreviated as the vortex state and the core region is called the vortex core, while the quadrant center-convergent topological domain structure will be abbreviated as the center state and the core region is called the center core. These domain structures are both topologically non-trivial and can be called as topological states too. More detailed analysis of PFM amplitude/phase images for identifying of vortex state and center state are presented in Supplementary (Section B and Fig. 2), wherein the PFM images recorded at 0º, 90º, also at 45° and 135° angles are included to further confirm their topologically non-trivial properties.

To illustrate the correlation between domain structure and conduction behavior, the two topological states were mapped in the conductive atomic force microscopy (CAFM) mode under a bias voltage of 2.0 V (see Fig. 1c & 1d, C-AFM). One can clearly identify the high conduction core regions for both states, along with the relatively lower conduction paths within the cross-shaped domain walls. For the vortex state, as shown in Fig. 1c, the conduction level at the core is ~ 3.0 nA, about three orders of magnitude larger than that of the 71° NDWs whose current level is only a few pA, hence the overall conductive pattern displays a highly conductive core plus four relatively faint cross-wings. The CAFM contrast of these cross-shaped NDWs is too faint to be visible in some cases, and thus the overall conduction pattern displays simply a bright spot at the core. However, for the center state, the feature is somewhat different. Besides the highly conductive core region, the CDWs also exhibit rather high conductive levels (~ 1.0 nA). As shown in Fig. 1d (C-AFM), the overall conduction pattern of the center state shows four bright wings from the conductive CDWs meeting at an even more conductive core (~ 3.0 nA). These unique features in the CAFM patterns constitute the unambiguous hallmarks for the two types of states, which provide an alternative way to identify these states. For instance, the highly conductive cross-wings in the center state are consistent with earlier observations [18], nonetheless the highest conducting core has not been previously reported.

More precise evaluation of the conduction level for the cores and domain walls of the two states can be seen from the current spatial-profiles (see Fig. 1e & f), extracted from the CAFM maps. It was found that conduction levels for different types of domain cores / walls ranking from the high to low levels are: center core (~ 3.5 nA) ≈ vortex core (~ 3.3 nA) > CDWs (1.3 nA) >> 71° NDWs (~ a few pA), given the identical probing bias of 2.0 V. These results can be further verified by the current-voltage (I-V) curves, obtained *via* placing a stationary tip on the specific core / wall regions and sweeping the tip bias between 0 and 4.0 V (as shown in Fig. 1g). It is also noted that the conduction level of vortex cores in this work is two orders of magnitude larger than previously reported values in artificially created vortex cores in BFO films [26], indicating a dissimilar conducting behavior. On the other hand, the conduction level of the center core is three-times larger than that of CDWs, implying a conductivity origin for the center core region different from that of the CDWs. Therefore, it is necessary to unveil the conduction mechanisms of these two types of topological cores. Moreover, the capability of deterministic tuning of these small topological objects is also a fundamental issue that deserves further attention.

**Temperature-dependent conduction behaviors**

To offer insights into the conduction mechanisms, the temperature-dependent conduction behaviors were probed (see Fig. 2). Fig. 2a & 2b are the CAFM images for the vortex state and center state within two nanoislands. A set of CAFM images were collected in the temperature range from 25 °C to 150 °C, and their corresponding PFM images at room temperature can be found in Supplementary Fig. 4. Clearly, the conduction levels of the two cores decrease gradually with increasing temperature, whereas the 71° NDW exhibits a monotonously increasing conduction. This trend is well-reproducible, confirmed in a number of nanoislands. It is also noted that the conductivities for both types of cores always return back to the initial levels after cooling-down the samples back to room temperature, indicating that these topological states are rather robust against heating.

From the CAFM mapping, we can extract the current profiles as shown in Fig. 2c & 2d, and plot the current-temperature (I-T) curves accordingly (see Fig. 2e). The I-T curves for both types of cores exhibit the negative temperature coefficients, which can fit to the well-known metallic conduction relation (I ~ $I_0(1 + a(T - T_0))^{-1}$), manifesting a metallic conducting behavior that is analogous to that of high conduction CDWs [8, 10]. In contrast, the NDWs exhibit positive temperature coefficients, which is a typical characteristic of semiconducting behavior.

To further comprehend the conduction behaviors of the two topological cores, we measured the local I-V curves at different T, as shown in Supplementary Fig. 5a & 5b. One can see distinctly different conduction behaviors in two different bias ranges. In the high bias range (V > 1.7 V), roughly liner I-V relations can be identified, implying a metallic behavior. In the low bias range (V < 1.7 V), the I-V curves exhibit nonlinear behaviors with a positive T-coefficient and conform to the Richardson-Schottky-Simmons emission model (Supplementary Fig. 5c ~ f), suggesting a thermionic emission (insulating) conducting behavior [8]. This claim can be also verified by the I-T curves measured at different bias voltages (Supplementary Fig. 5g & 5h), whereby the slopes of the I-T curves gradually shift from the positive value to negative value with increasing bias voltage. A metal-insulator transition for both types of cores occurs at a bias threshold of ~ 1.7 V.

The observed phenomena at the center core can be understood from the band structure variation within the core, as schematically illustrated in Supplementary Fig. 6. At the head-to-head center core, there exists a large amount of uncompensated bound-charge which tends to attract a high density of electron carriers and charged point defects (e.g. vacancies) to balance the bound-charge. This results in a dramatic band-bending inside the conducting core which lowers the conduction band to be below the Fermi level, forming the ultra-small highly conductive 1D metallic channel at the core, as an analogue to the conduction behavior in the CDWs [8, 10, 11]. When the metallic channel contacts with the electrode, there may occur a very narrow insulating gap or an unconducive domain region [11] close to the electrode, giving rise to the Schottky barriers between the metallic core and electrodes. Therefore, the system exhibits a thermionic emission behavior in the lower bias range.

At sufficiently large electric bias, the unconducive domain region can be annihilated, and the large electric field significantly narrows the tunneling barrier from the insulating gap, which greatly reduces the interfacial resistance. As a result, the metallic conduction behavior of the 1D channel becomes dominant over the interfacial thermionic effect. This is consistent with the observation of a threshold in the I-V curves for both types of cores, in which the nonlinear (non-Ohmic) behavior dominates below a bias voltage of 1.7 V yet linear dependence is shown at higher bias voltage (see Supplementary Fig. 5a & 5b). It is also noteworthy that the vortex core does not contain such uncompensated bound charge. Why such vortex core region still exhibits metallic conduction, in analogue to the center core, remains an open question.

**Phase-field simulation of conduction behaviors**

To understand why these two different topological cores exhibit roughly similar conduction behaviors, we conducted phase-field simulation. Here, the nanoislands are described as nanoscale cylinders (the simulation details are shown in the Method section). Although the size of the model cylinders (70 nm in diameter) is smaller than the real nanoislands, nonetheless it is sufficient to study the different conduction behaviors of the two types of ultra-small cores.

Figure 3a to 3f show the domain structures and conductivity contours of three center states (three columns) derived from a phase-field simulation. We first consider a perfect center domain structure without any distortion (see Fig. 3a & 3d), which produces a clear cross-shaped conduction pattern consisting of four CDWs along with one small while highly conductive core, in good agreement with the experimentally observed hallmark conductive pattern (see Fig. 1d). After further relaxation of the domain structure, these four CDWs may become severely distorted (*e.g.* zigzag-shaped), leading to an apparent non-uniformity and partial loss of the wall conductivity, while the core remains highly conductive (see Fig. 3b & 3e). Furthermore, given a direct relaxation from an initially random polarization state under electric fields produced by the bias near a PFM tip, in order to mimic our experimental situation, one may see bending or tilting in the CDWs to some extent, and also the non-uniform conductivity as well, noting that the high conductivity of the core region is well preserved (see Fig. 3c & 3f).

It is suggested that the local wall distortions (e.g. bending or tilting) can greatly redistribute the bound charges and electrostatic energy, hence sizably modulating the wall conductivity [19]. The formation of the zigzag-shaped walls rather than straight ones is driven by the release of large electrostatic energy, explaining the apparent variation and non-uniformity of the wall conductivity. In contrast, the conductivity of the center core region is rather resilient to local distortions or disturbance, manifesting the nature of topological protection. Therefore, the stable and high conduction of the center core is an intrinsic property of topological protection, an advantageous merit for device applications.

Very differently and also unexpectedly, the simulation shows that the vortex core exhibits very low conductivity (only a rather small enhancement) at zero bias field (see Fig. 3g & 3j), which contradicts with our experimental result. To clarify this discrepancy, we can apply a scanning bias (1.0 V) at the core to mimic the real current reading situation, and it turns out that the conductivity of vortex core can be significantly enhanced *via* forming a twisted state (see Fig. 3h & 3k, and see the bright dot at the center shown in Fig. 3k). This twisted state is somehow akin to a highly conductive head-to-head center core, as further supported by the observation of severe twisted core induced by 2.0 V bias (see Fig. 3i & 3l). Such severely twisted core is very close to a convergent center domain core, while a flux-closure domain pattern is still preserved outside the core region. The twisted core was indeed observed experimentally, whereby a very large bias voltage (7.0 V) was used to stabilize it for a short while (see Supplementary Fig. 7). This is probably due to that both vortex and center states share the same winding number (+1), making it easier to convert the vortex core to a twisted state at nanometer scale. The aforementioned phenomena can well explain why the vortex core exhibits such similar conduction behavior to that of the center core, because the measured conductivity is produced from the induced twisted vortex core stabilized during the reading process when the instant domain structure around the core was rather similar to a center core.

It is noted that slightly enhanced conductivity (with current of a few pA at a reading bias of ~ 2.0 V) at an artificial created vortex core in BFO film was already previously reported, which was also interpreted by the occurrence of a metastable twisted structure that contains conductive CDW segments [26]. To monitor the enhanced conductivity of the vortex core by CAFM, one must pre-write the metastable twisted structure by a local bias, and the observed conductivity enhancement is rather unstable during reading process. In contrast, the twisted vortex core in this work exhibits significantly enhanced conductivity (~ 3.0 nA at 2.0 V), which can be directly induced during the CAFM scan (at scan bias of 2.0 V) without a pre-written process. Such conductivity enhancement behavior is reproducible in different vortex cores and rather stable, likely a universal property for certain topological defects.

One can also estimate the diameter of the high conduction center/vortex cores from the full-width of half-maximal of the simulated conductivity maps, and it is very small with the lateral size < 2.5 nm (see Supplementary Fig. 8). This scale is close to the lateral size of a typical CDW [8, 18], giving rise to an ultrahigh current density of ~ $10^4$ A/cm$^2$. Such a superfine and highly conductive 1D metallic channel can be considered as a kind of quasi-1D electron gas (q1DEG), somewhat analogous to the quasi-two-dimensional electron gas (q2DEG) frequently observed in certain CDWs [8, 10, 11], while it is out of the scope of this work.

**Controllable creation and elimination of the high conduction topological cores**

Certainly it is imperative to discuss the possible applications of these conductive topological cores in configurable devices. A fundamental issue is to achieve a controllable manipulation of these conductive topological cores from one state to the other. For this motivation, we performed extensive investigations and successfully achieved controllable and reversible creation and elimination of the vortex and center states separately for our nanoislands, *via* applying a suitable scanning bias with the conductive AFM tip (see Supplementary section G and Fig. 9).

The creation and deletion of the conduction channels can be clearly seen from the evolution of conduction patterns for two different topological states (corresponding PFM data can be found in Supplementary Fig. 10) in a nanoisland array. As shown in Fig. 4a, the initial pristine state usually exhibits very low conductivity (~ 1.0 pA) from the wedge-like domain pattern with a net downward out-of-plane polarization component (abbreviated as *wedge domain state*). Upon applying a scanning bias of + 5.5 V on the whole array, all the nanoislands exhibit the bright cross-shaped conduction pattern which is a typical hallmark of center domains (with upward out-of-plane polarization). After applying a negative bias of - 3.5 V on four selected nanoislands (marked with red and blue circles), the selected center states switch back to the low conduction states (wedge domain state). Further application of a bias voltage of + 5.5 V on two of the previously selected nanoislands creates two center states (in blue circles), and the application of a bias voltage of + 3.5V on the other two nanoislands creates two vortex states (with upward polarization, in red circles) as reflected by the characteristic conduction patterns that show a single high-conduction core only. This suggests the capability of reversibly writing and deleting individual topological states as well as their conductive cores.

It is also noted that a vortex state can be switched to a center state by applying a bias of 5.5 V, nonetheless this switching is not reversible (see Supplementary Fig. 9e & 9f). Therefore, a scheme by applying a suitable bias voltage allows a programmable writing of different topological states individually as well as a programmable control of their conduction states. The topological domain switching process, induced by the tip-bias, is schematically summarized by the triangle scheme in Fig. 4b, illustrating the capability of reversible creation/elimination of vortex and center states inside a single nanoisland, as well as the topological switching from a vortex state to a center state.

The capability of controllable creation/elimination of these conductive states holds promise for their implementation in high-density memory device. A schematic of the exemplified conceptual crossbar memory is shown Fig. 4c, by exploiting the programmable topological states (either vortex or center state) as storage units which enable the non-destructive readout through the corresponding core states. We have tested the write/read performance for such a randomly selected device, as shown in Fig. 4d & 4e. It is found that the resistance switching between the low conduction wedge domain state and high conduction vortex/center states produces an on/off resistance ratio larger than $10^3$ which is rather stable against 50 switching cycles without apparent fatigue (see Fig. 4d), while this cycle number can be much larger. The on/off resistance states can also be maintained stable up to $1.0 \times 10^7$ s (17000 minutes, see Fig. 4e). The high stability of the device is mostly due to the topological protection nature of the conductive cores, which enables good resilience against local disturbances or thermal fluctuations, a remarkable advantage over 2D domain walls whose local conductivity can be sizably influenced by local disturbance [19].

Although the center core exhibits a similar conduction level to that of the vortex core, the overall conduction state of a center state (including both core and CDWs) is indeed much larger compare to a vortex state, due to the contribution of much higher conductivity in the elongated CDWs. Therefore, the switching among the wedge state, vortex state, and center state also has the potential to be utilized to develop multi-level memory devices, which deserves further investigations.

We would like to point out that the superfine dimension of these core conduction channels also offers an excellent possibility for scaling the device dimension down to sub-3 nm, given the sufficient capability of non-destructive current read-out from the conductive topological cores. Specifically, it was predicted that the center domain state can be stabilized at a small dimension of 16 nm [18], yet vortex can reach a size even as small as ~ 3 nm [35]. Besides, these qusi-1D cores/topological defects are confined in nanoisland structures, which is also compatible to modern semiconductor high density integration processes.

These beneficial features create a potential pathway towards the low-energy consumption, high-efficient, stable, ultra-dense, and configurable electronics devices. Specifically, the ultra-small (~ 3.0 nm) vortex core poses a possibility of developing non-volatile memories with an areal density of 60 Tb per square inch [35], around four orders of magnitude higher than that of modern random-access memories. Moreover, the finding of these fascinating properties in superfine topological objects might inspire future efforts to seek other exciting unexplored properties and associated application potentials, for instance, by exploring their responses to external stimuli, e.g. strain, electric field, magnetic field, and light illumination, which might eventually enrich the field of topological electronics.

**Conclusion**

Two types of 1D topological defects, namely the vortex cores and center cores, in high quality BFO nanoislands exhibit highly conductive and metallic behaviors and behave like quasi-1D metallic conduction channels. The enhanced conductance of the center cores is an intrinsic property associated with charge accumulation whereas the high conductance of the vortex cores arises from the electric field-induced twisted vortex core structure. These conductive channels can be reversibly created and deleted, producing resistance switching with an on/off ratio larger than $10^3$. This electro-resistance functionality of these ultra-small 1D topological objects are stable over many cycles of switching and have a long retention time, thus it can potentially be applied to high performance programmable topological electronic devices, e.g. ultrahigh density non-volatile memory with none-destructive current readout.

ferroelectric domain walls. *Nat. Nanotechnol.* **13,** 947–952 (2018).

31. Li, S.-Z. et al. Topological defects as relics of emergent continuous symmetry and Higgs condensation of disorder in ferroelectrics. *Nat. Phys.* **10,** 970–977 (2014).

32. Sharma, P. et al. Nonvolatile ferroelectric domain wall memory. *Sci. Adv.* **3,** e1700512 (2017).

33. Jiang, J. et al. Temporary formation of highly conducting domain walls for non-destructive read-out of ferroelectric domain-wall resistance switching memories. *Nat. Mater.* **17,** 49−56 (2017).

34. Vasudevan, R. K. et al. Domain wall geometry controls conduction in ferroelectrics. *Nano Lett* **12,** 5524 (2012).

35. Naumov, I. et al. Unusual phase transitions in ferroelectric nanodisks and nanorods. *Nature* **432,** 737–740 (2004).

36. Kalinin, S. et al. Vector piezoresponse force microscopy. *Microsc. Microanal.* **12,** 206–220 (2006).

**Acknowledgments**

The authors would like to acknowledge the financial support from the National Key Research and Development Programs of China (Grant Nos. 2016YFA0201002, 2016YFA0300101), the National Natural Science Foundation of China (Grant Nos. 11674108, 11834002, 51272078, 51721001), the Project for Guangdong Province Universities and Colleges Pearl River Scholar Funded Scheme (2014), the project for Basic and Applied Basic
20

research Foundation of Guangdong Province (No.2019A1515110707), the Natural Science Foundation of Guangdong Province (No. 2016A030308019), and the Science and Technology Planning Project of Guangdong Province (No. 2015B090927006). Y. Z gratefully acknowledges the financial support from China Scholarship Council (No. 201706190099). The work at Penn State is supported by the US National Science Foundation under grant number DMR-1744213.

**Author contributions**

X. S. Gao conceived and designed the experiments. W.D. Yang and G. Tian conducted the main experiments. G. Tian and C. Chen contributed to the sample fabrication and XRD measurements. W. D. Yang carried out the SEM, AFM, PFM CAFM and KPFM measurements. Y. Zhang, F. Xue, and L.-Q. Chen contribute to the phase field simulations. D. Y. Chen, Z. Fan, Z. P. Hou, J. W. Gao contributed to the data interpretation. X. S. Gao, L.-Q. Chen, and J.-M. Liu conducted the data interpretation and co-wrote the article. All authors discussed the results and commented on the manuscript.

**Competing interests**

The authors declare no competing financial interest.

**Additional Information**

**Supporting Information** is available free of charge at XXXXXXXXXXX.

**Methods**

**Fabrication of nanodot arrays.** The fabrication procedure for the nanoisland arrays have been illustrated in the schematic flowchart in Supplementary Fig. S1, based on nano-sphere patterning on well-epitaxial BFO thin films. Firstly, a 35 nm-thick epitaxial $BiFeO_3$ thin film and a ~20 nm-thick epitaxial $SrRuO_3$ bottom electrode layer were deposited on the (100)-oriented $SrTiO_3$ substrates by pulsed laser deposition (PLD). Then, the polystyrene spheres (PS) pre-dispersed in a mixture of ethanol and water were transferred onto the BFO film, to form a close-packed monolayer. The sizes of the nanospheres were then shrunken by plasma etching to form a discrete ordered island array, which was followed by $Ar^+$ ion beam etching with appropriate durations. Finally, the PS template was removed by chloroformic solution and finally the periodically ordered BFO nanoisland arrays were obtained. After the patterning, the samples were also annealed at oxygen ambiance at 400 ºC.

**Microstructural characterizations.** The structures of nanoislands were characterized by X-ray diffraction (PANalytical X′Pert PRO), including θ-2θ scanning and reciprocal space mapping (RSM) along the (103) diffraction spot. The top view surface images were obtained by scanning electron microscopy (SEM, Zeise Ultra 55), and the topography images were taken by atomic force microscopy (Asylum Cypher AFM).

**PFM and CAFM characterizations.** The ferroelectric domain structures of the nanoislands were characterized by piezoresponse force microscopy (PFM) with a scanning probe mode (Cypher, Asylum Research) using conductive PFM probes (Arrow EFM, Nanoworld). The local piezoresponse loop measurements were carried out by fixing the PFM probe on a selected nanoisland and then applying a triangle square waveform accompany with a small *ac* driven voltage from the probe. Using vector PFM mode, one can simultaneously map the vertical (out-of-plane) and lateral (in-plane) piezoresponse signals from the nanoisland one by one. To determine the domain structures, both the vertical and lateral PFM images were record at different sample rotation angles. For this, we marked the sample before the rotations, so that the same scanned area can be tracked in different scan. The conductive current distribution maps, current-voltage (I-V) measurement were characterized by conduct-tip atomic force microscopy (CAFM) by using conduct probes (CONTV-PT, Bruker).

**Phase-field simulation:** In the phase-field model, we consider both the polarization vector $P_i$ ($i$ = 1, 2, 3) and the oxygen octahedral tilt vector $\theta_i$ ($i$ = 1, 2, 3) as order parameters to simulate the domain patterns in BFO nano-islands [37]. The total Helmholtz free energy of BFO includes the bulk, gradient, elastic, and electrostatic free energy terms which can be written as [37-39]:

$$F = \int_V \begin{bmatrix} \alpha_{ij} P_i P_j + \alpha_{ijkl} P_i P_j P_k P_l + \beta_{ij}\theta_i\theta_j + t_{ijkl} P_i P_j \theta_k \theta_l + \\ + \frac{1}{2} g_{ijkl} \frac{\partial P_i}{\partial x_j}\frac{\partial P_k}{\partial x_l} + \frac{1}{2}\kappa_{ijkl}\frac{\partial \theta_i}{\partial x_j}\frac{\partial \theta_k}{\partial x_l} + \\ \frac{1}{2} c_{ijkl} \left(\varepsilon_{ij} - \varepsilon_{ij}^0\right)\left(\varepsilon_{kl} - \varepsilon_{kl}^0\right) - E_i P_i - \frac{1}{2}\varepsilon_0 \kappa_b E_i E_j \end{bmatrix} dV, \qquad (1)$$

where $\alpha_{ij}$, $\alpha_{ijkl}$, $\beta_{ij}$, $\beta_{ijkl}$, and $t_{ijkl}$ are the Landau polynomial coefficients. $g_{ijkl}$ and $\kappa_{ijkl}$ are the gradient energy coefficients for $P_i$ and $\theta_i$, respectively, with $x_i$ the spatial coordinate. $c_{ijkl}$ is the elastic stiffness tensor, $\varepsilon_{ij}$ is the total strain, and $\varepsilon_{ij}^0 = h_{ijkl} P_k P_l + \lambda_{ijkl} \theta_k \theta_l$ is the eigenstrain with $h_{ijkl}$ and $\lambda_{ijkl}$ the coupling coefficients. $E_i = -\partial \varphi / \partial x_i$ is the electric field with $\varphi$ the electrostatic potential, $\varepsilon_0$ is the permittivity of vacuum, and $\kappa_b$ is the background relative dielectric constant.

All the values of coefficients can be found in previous literatures [37].

The temporal evolution of order parameters is simulated by the time-dependent Ginzburg-Landau equations $\partial P_i / \partial t = - L_P(\delta F / \delta P_i)$ and $\partial \theta_i / \partial t = - L_\theta(\delta F / \delta \theta_i)$ using the semi-implicit Fourier spectral method [40], where $L_P$ and $L_\theta$ are kinetic coefficients. For each time step, the elastic and electric driving forces can be calculated by solving the mechanical equilibrium equations $\sigma_{ij,j}=0$ and the electrostatic equilibrium equation $D_{i,i}=0$, where $\sigma_{ij}$ is the local stress and $D_i$ is the electric displacement. The spectral iterative perturbation method [41] is adopted.

The whole system grid is $256\Delta x \times 256\Delta x \times 32\Delta x$ with $\Delta x = 1.0$ nm. The system consists three parts, i.e., $256\Delta x \times 256\Delta x \times 14\Delta x$ for the substrate, $(70\Delta x)^2\pi \times 14\Delta x$ for the BFO circular island, and the rest for air. In the air and substrate, $P_i = 0$ and $\theta_i = 0$. The elastic stiffness in the substrate is assumed the same as BFO, while the elastic stiffness in air is zero. The electric boundary conditions of the BFO islands are short-circuit for the up and bottom surfaces and open-circuit for island surroundings. The electric potential of bottom interface $\varphi^{bot}$ is always zero, whereas the potential of top surface $\varphi^{top}$ is uniform with certain values or non-uniform induced by a PFM tip. For the latter, it is approximated by a Lorentz distribution [26]

$$\phi^{top}(x_1, x_2) = \phi_0 \left( \frac{\gamma^2}{r^2 + \gamma^2} \right), \qquad (2)$$

where $\varphi_0$ is the electric bias applied on the PFM tip, $r$ is the lateral distant from the site $(x_1,x_2)$ to the position of PFM tip, and $\gamma=15$ nm is the half-width at half-maximum of the tip.

The electrical conductivity of BFO nanoislands can be approximated from the electrostatic potential $\varphi$. If we assume that the negative charge carriers are dominant in BFO, the local conductivity can be estimated according to Boltzmann statistics as [42]

$$\sigma = N_0 e\mu \cdot \exp(-\frac{e\phi}{k_B T}), \tag{3}$$

where $N_0$ is the background carrier density, $e$ is the charge of an electron, µ is the carrier mobility, $k_B$ is Boltzmann constant, and $T$ is absolute temperature. To make the distributions of conductivity under different conditions comparable, we choose short-circuit boundary conditions with $\varphi^{top} = \varphi^{bot} = 0$ for certain domain structures when we calculate the conductivity.

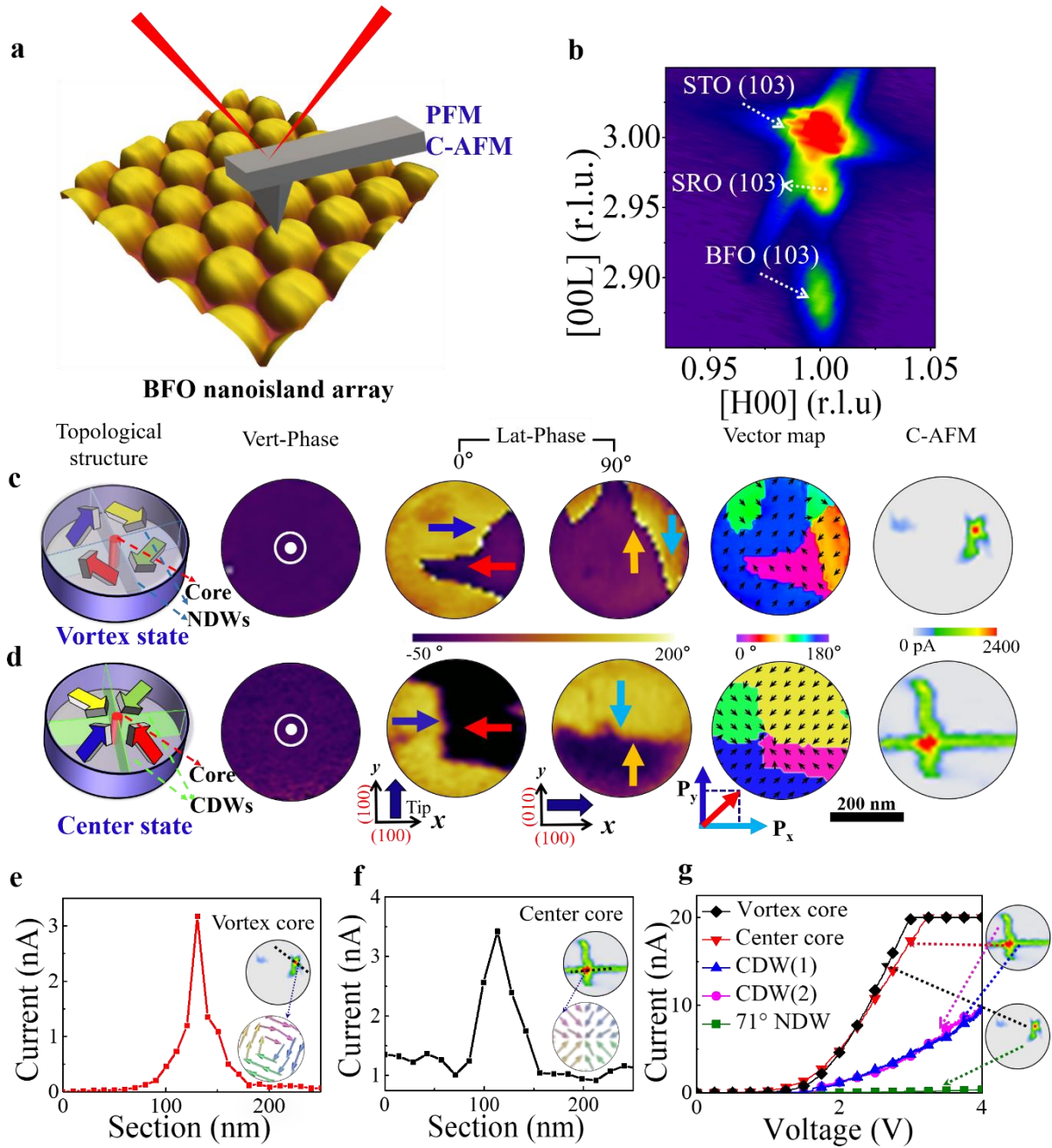

**Fig. 1 | Structures of BFO nanoslainds, as well as the domain structures and correspondent conduction patterns for both quadrant vortex and quadrant center topological states confined in two nanoislands. a,b,** The AFM topographical 3D image of an array of BFO nanoisland (a), and the RSM map for the nanoislands. **c,d,** The domain states and their corresponding C-AFM maps for two topological states: vortex

(c) and center (d) states. The micrographs (in **c** and **d**) from the left to the right are: schematic domain structures for typical vortex/center states in this work (with enhanced conductivity in topological cores and domain walls), the PFM vertical phase images illustrating the uniform upward out-of-plane polarization components for both nanoislands, the PFM lateral phase images recorded at sample rotation of 0º and 90º to evaluate the directions of in-plane polarization components respectively along *x-axis* (*[100]-axis*) and *y-axis* (*[100]-axis*), the in-plane polarization vector maps derived from the lateral PFM data, and corresponding CAFM maps. The arrows inside the PFM images present the in-plane directions of the polarization components perpendicular to the directions of PFM cantilever. **CDW** and **NDW** present charged domain wall and neutral domain wall, respectively. **e,f,** Extracted current spatial profiles from the CAFM maps for both vortex (e) and center (f) cores, extracted from **c** and **d**, respectively. The inserts in the figures show the schematic diagrams of local polarization configurations of the two types of topological cores. **g,** Temperature dependent conductive current (**I-V**) curves for the both topological cores and domain walls.

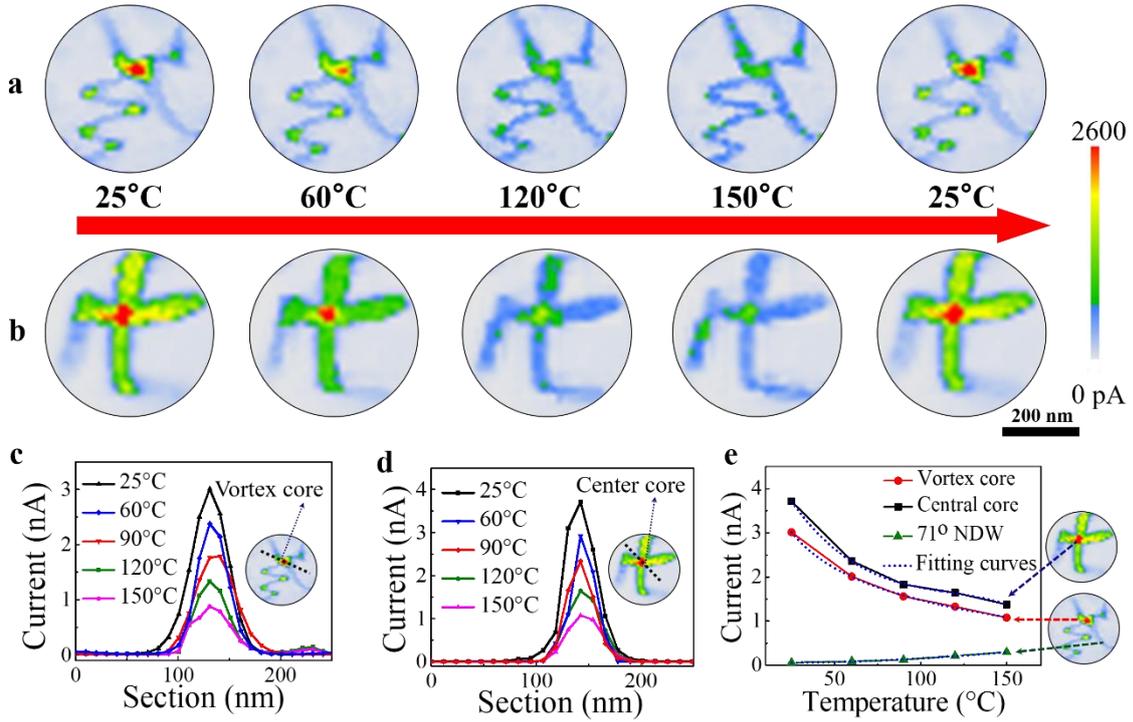

**Fig. 2 | Temperature dependent conductive behaviors for the vortex and center cores. a,b,** The CAFM maps at different temperatures for two presentative nanoislands, respectively contenting a vortex state (**a**) and a center state (**b**), along with some domain walls. The long red arrow (between **a** and **b**) presents the heating and cooling consequences. **c,d,** Extracted current spatial profiles from the CAFM image in **a** and **b** as a function of temperature for the two topological cores. **e,** Temperature dependent conductive current (*I-T*) curves for both topological cores and a NDW wall (for comparison purpose), wherein the *I-T* curves for the topological cores can fit well to the metallic conductance relation (I ~ 1(1 + a(T − $T_0$))$^{-1}$), while that of NDW conforms the thermal activation relation (I ~ exp($\Delta E_v$/kT).

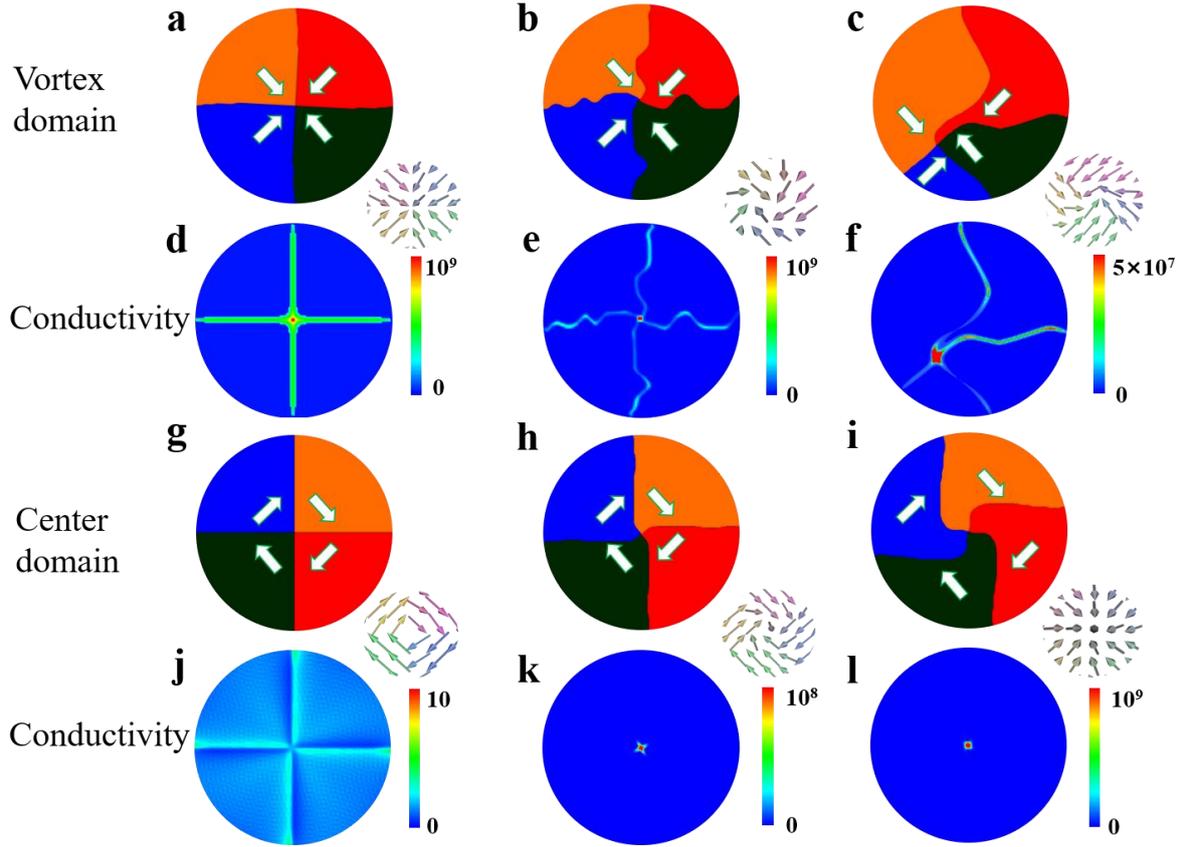

**Fig. 3 | Phase-field simulation of the conduction states for both vortex and center states. a-f,** Simulated domain structures and corresponding conductivity contours for the quadrant center states: perfect center domain state (a and d), distorted center state with zigzag shape charged domain walls generated *via* relaxing from a perfect center state (b and e), distorted center state with curved charged domain walls formed *via* relaxing from an initial random polarization state (c and f). **g-l**, Simulated domain structures and corresponding conductivity contours for the quadrant vortex states: perfect quadrant vortex state (g and j), distorted vortex state with a twisted core induced by applying a scanning bias of 1 V (h and k), and distorted vortex state with a severe twisted core induced by a scanning bias of 2V (i and l). The local polarization configurations adjacent to topological core regions were also enlarged and inset besides the topological domain structures.

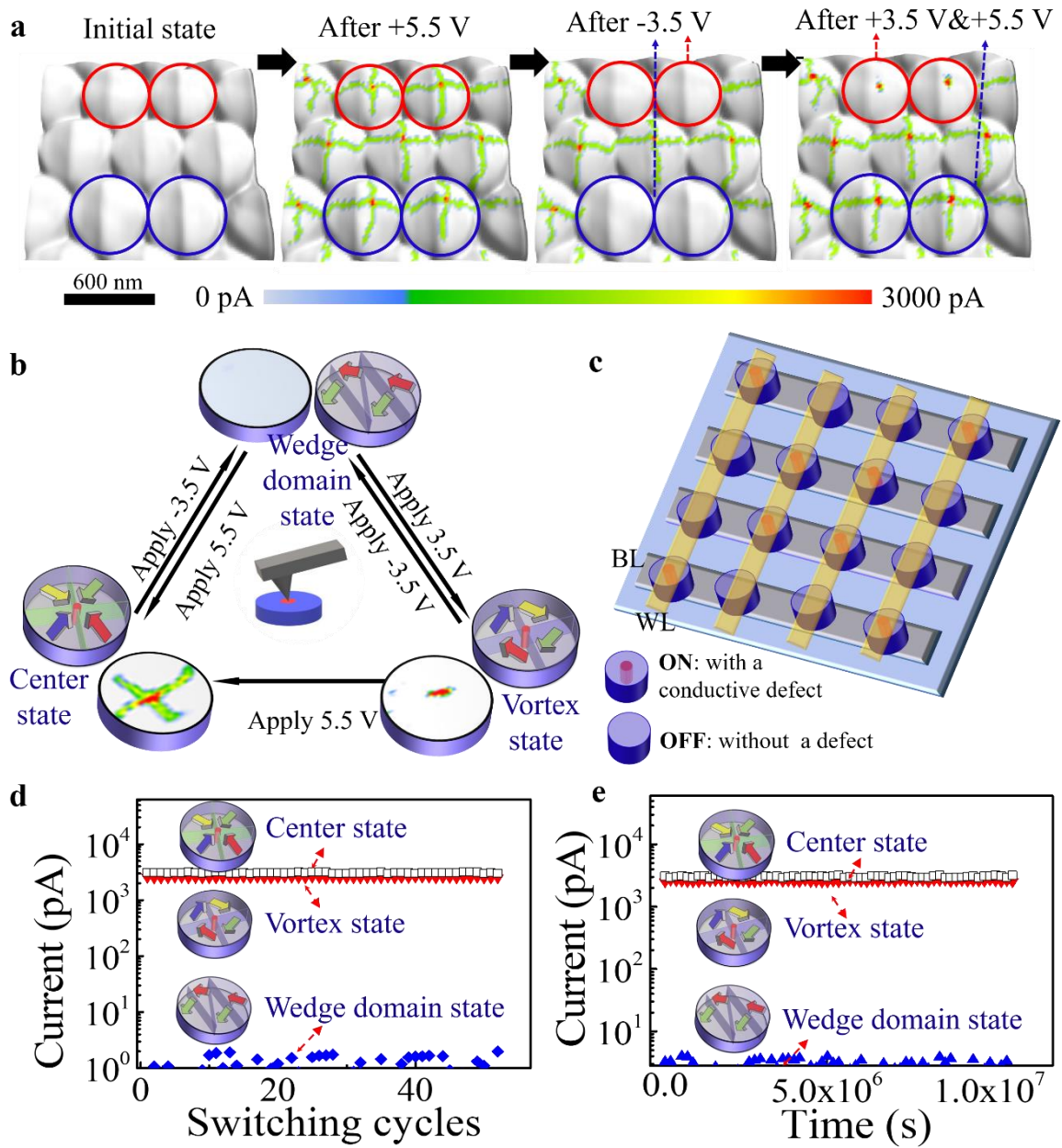

**Fig. 4 | Manipulation of conductive states *via* controlled creation and elimination of vortex/center topological states. a,** Control creation/elimination of topological states in selected nanoislands, as reflected by the hallmarks of conductive states, from left to right are: initial low conduction state with wedge domains (abbreviated as *wedge domain states*), creation of center states over the whole array by applying a scanning bias voltage of +5.5 V, elimination of center states to form low conductive wedge domain states in four

selected nanoislands by applying -3.5 V, and creation of two vortex states and two center states by applying +3.5V and +5.5 V, respectively, in the four previously selected nanoislands. **b,** Schematic diagram to summarize the creation/elimination and switching of topological states, as well as the evolution of corresponding conduction states. **c,** A schematic conceptual cross-bar memory device using the programable metallic conduction channels in topological defects (either vortex or center cores) as data bits. **d,e,** The repeatability (d) and retention properties (e) of resistance changes between low and high conductance states *via* writing and erasing of topological cores, which exhibit a large on/off resistance ratio over $10^3$, and can maintain stable over 50 switching cycles and retention time of $10^7$ s, manifesting potentially high performance for the nonvolatile memory devices based on these functional topological defects.

**Supplementary Information**

**Quasi-one-dimensional metallic conduction channels in exotic ferroelectric topological defects**

Wenda Yang,[†a] Guo Tian,[†a] Yang Zhang,[‡] Fei Xue,[‡] Dongfeng Zheng,[†] Luyong Zhang, Yadong Wang,[†] Chao Chen,[†] Zhen Fan,[†] Zhipeng Hou,[†] Deyang Chen,[†] Jinwei Gao,[†] Min Zeng,[†] Minghui Qin,[†] Long-Qing Chen,[‡] Xingsen Gao,[*,†] and Jun-Ming Liu[†§]

[†] Guangdong Provincial Key Laboratory of Quantum Engineering and Quantum Materials and Institute for Advanced Materials, South China Academy of Advanced Optoelectronics, South China Normal University, Guangzhou 510006, China

[‡] Department of Materials Science and Engineering, The Pennsylvania State University, University Park, Pennsylvania 16802, United States of America

[§] Laboratory of Solid State Microstructures and Innovation Center of Advanced Microstructures, Nanjing University, Nanjing 210093, China

[*]Corresponding author: xingsengao@scnu.edu.cn

[a]These authors contributed equally

## A. Fabrication of epitaxial BiFeO$_3$ nanoislands

The fabrication procedure for the BiFeO$_3$ (BFO) nanoisland arrays is shown in the schematic flowchart of Supplementary Fig. 1a, and it was developed based on a nano-sphere patterning scheme on well-epitaxial BFO thin film [1, 2]. In brief, the epitaxial BFO thin film of ~ 35 nm in thickness, along with a ~ 20 nm thick epitaxial SrRuO$_3$ (SRO) layer, was deposited on the (100)-oriented SrTiO$_3$ substrate by pulsed laser deposition (PLD) using a KrF excimer laser (wavelength $\lambda$ = 248 nm) at 680 °C in oxygen ambient of 15 Pa, with a laser pulse energy of 300 mJ and a repetition rate of 8 Hz. Subsequently, the polystyrene spheres (PS) dispersed in a mixture of ethanol and water were transferred onto the as-grown BFO thin film surface to form a close-packed monolayer. The nano-sphere monolayer was then etched to the desired size by oxygen plasma to form a discrete ordered array. This procedure was followed by Ar$^+$ ion beam etching with appropriate etching time. Finally, the PS template was lift-off by chloroformic solution and a periodically ordered BFO nanoisland array was obtained, completing the patterning process. After that, the sample with nanoisland array was annealed at oxygen ambiance and at temperature of 400 °C for 30 min to reduce the defects and residual strain.

**Ion beam etching parameters.** The Ar$^+$ ion beam etching was conducted in a vacuum pressure of $8.0 \times 10^{-4}$ Pa at room temperature. During the etching, the incident ion beam was alighted in perpendicular to the sample surface, while the etching parameters were carefully optimized, giving rise to a cathode current of 15.7 A, an anode voltage of 50 V, a plate voltage of 300 V, an ion accelerating voltage of 250 V, a neutralization current of 13 A, and a bias current of 1.2 A.

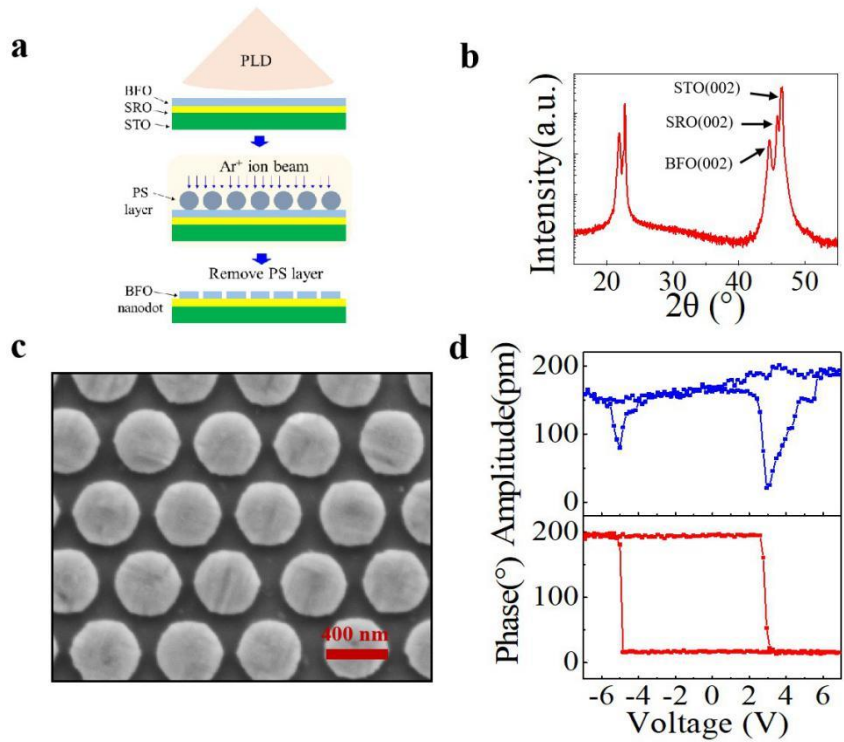

**Supplementary Figure 1 | The structural and piezoelectric characterizations of a BFO nanoisland array. a,** A schematic flowchart for the fabrication process. **b-c**, The XRD θ-2θ scan (b), and the SEM image (c) of the nanoisland array. **d**, The local piezoresponse butterfly-like amplitude-voltage hysteresis loop and phase-voltage loop on a randomly selected nanoisland.

## B. Domain structure reconstruction

A major progress to achieve this work has been the reconstruction of domain structures in these nanoislands using the piezoresponse force microscopy (PFM) data. This PFM microscopy allows the simultaneous mapping of vertical (out-of-plane) and lateral (in-plane) amplitudes (abbreviated as V-amp and L-amp) and phases (abbreviated as V-pha and L-pha). In order to determine the local polarization directions, one can combine the PFM data recorded at different cantilever orientations. In most cases, we chose the PFM imaging data at the cantilever orientation of 0° and 90°, with respect to the sample surface, and thus the polarization components along the defined *x*-axis and *y*-axis ((100)-axis and (010)-axis of BFO in our case, respectively), were determined [3, 4].

Details of the analysis on the two types of domains: the vortex domain and central domain, are illustrated sequentially in Supplementary Fig. 2. Given the measured PFM imaging data for two selected nanoislands upon the clockwise rotation for 0°, 45°, 90°, and 135° respectively, we connected the image contrast with the three-dimensional polarization configuration. The different contrasts in the lateral-phase image (L-pha) represent the two opposite in-plane directions, where the in-plane projections of the polarization are perpendicular to that of the cantilever. The dark lines in the amplitude image (L-amp) mark the domain wall where the piezoelectric response should be weak. Based on these PFM images, one can conclude that the in-plane projection of the polarization in either of the two types of nanoislands contains four quadrant domains. Based on the 0° and 90º measured data (which determine the directions of polarization components along [100]-axis and [010]-axis of BFO lattice, respectively) and also given the fact that the in-plane projections of polarization are aligned along the four [110] directions, we can conclude that the in-plane projections must develop into four quadrant domains. From the uniform dark contrasts in vertical PFM images, one can determine that the vertical components of polarization are aligned upwards for both nanoislands.

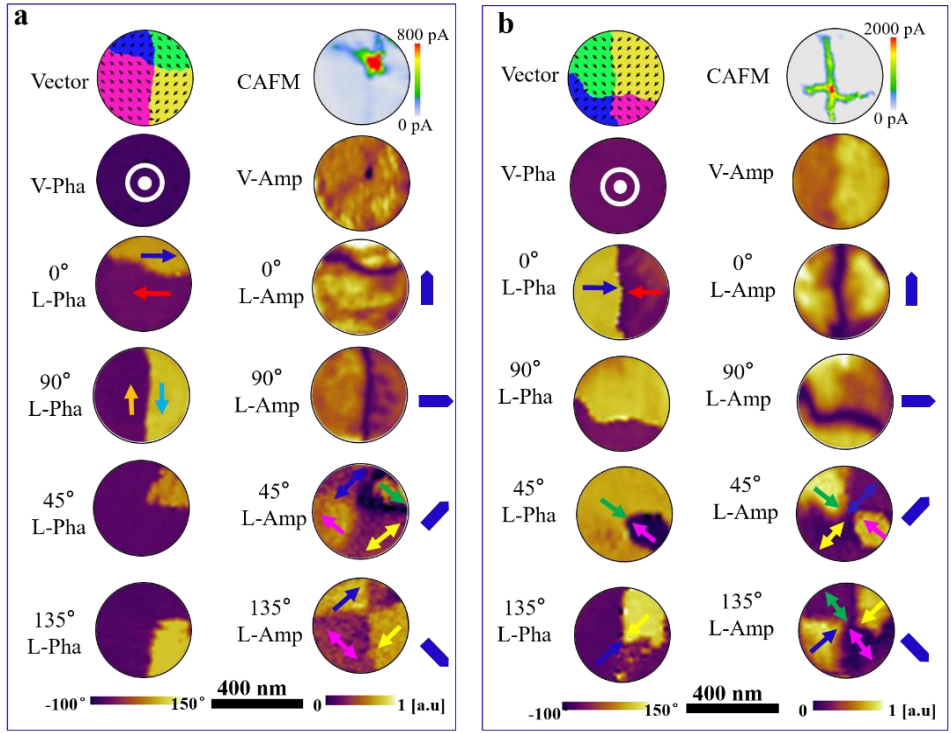

**Supplementary Figure 2 | The domain structure reconstruction and corresponding conduction maps for the vortex domain and center domain. a,b,** A collection of the in-plane polarization vector maps and their corresponding CAFM maps, vertical PFM phase (V-Pha) and amplitude (V-Amp) images, lateral PFM phase (L-Pha) and amplitude (L-Amp) images, captured at various sample-rotation angles (0º, 90º, 45º, 135º), for both the vortex state (a) and center state (b). The big blue arrows outside the images present the directions of cantilever, and the small/narrow arrows inside the images present the local in-plane directions of polarization components perpendicular to the cantilever. The double-head arrows present the undefined directions of local polarization components which are parallel to the directions of cantilever, noting that the marked regions show dark contrast in lateral PFM-amplitude image.

The first nanoisland, as shown in Supplementary Fig. 2a, is characterized by meeting of four 71º head-to-tail neutral domain walls (71° NDWs) where the four domains have the upward out-of-plane polarization components, forming a flux-closure pattern. This is a typical feature for a quadrant vortex state. In the second nanoisland, as shown in Supplementary Fig. 2b, there exist four head-to-head charge domain walls (CDWs) joining at a core, along with four quadrant domains having the upward out-of-plane polarization components too and each in-plane polarization component pointing inward towards the core, a typical characteristic of center domain state.

It should be mentioned that the assigned domain structure also fits the contrast evolution for the 45º and 135º rotations respectively. Besides, the CAFM images of the two nanoislands match well with their domain structure, where the domain walls show a marked enhancement in the conduction current with respect to the domain interior regions.

## C. Creation of vortex and center topological states

Given the reliable reconstruction of domain structure, it is now possible to implement the creation and removal of the two types of topological domain structures. We start from the initial domain structure which is a wedge-like pattern (abbreviated as wedge domain state). The vortex and center states can be created by scanning the bias voltages, given different writing schemes. For example, a bias of ~ 3.5 V can create a vortex domain state while a bias of ~ 5.5 V allows writing a center domain, as shown in Supplementary Fig. 3. These PFM images can be used for domain structure reconstruction and the obtained PFM lateral phase images at angles of 0º and 90º are presented, where the directions of local polarization components along the *x*-axis and *y*-axis of BFO can be determined, together with the in-plane local polarization vector maps and the CAFM maps.

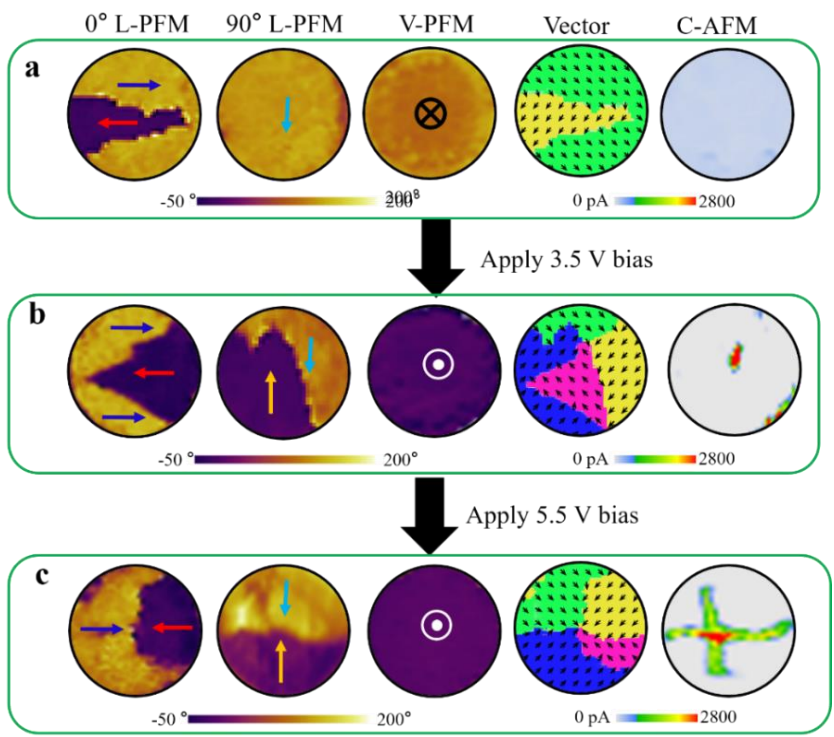

**Supplementary Figure 3 | The creation of topological vortex and center states. a-c**, The domain structures and corresponding CAFM patterns for: the initial wedge domain state with low conduction (a), and the vortex state created by a 3.5 V biased scanning (b), and the center state obtained after a 5.5 V biased scanning (c). The current level for the neutral domain walls is too weak to be visible in the images for both the initial wedge domain state and vortex state. The micrographs from the left to the right are: the PFM lateral phase images (0º and 90º) used

to evaluate the directions of local polarization components (along *x*-axis and *y*-axis respectively), the vertical phase images, the in-plane local polarization vector maps, and the CAFM maps. The small arrows inside the images label the in-plane directions of local polarization component perpendicular to the directions of cantilever. The diameter for each nanoisland is 400 nm.

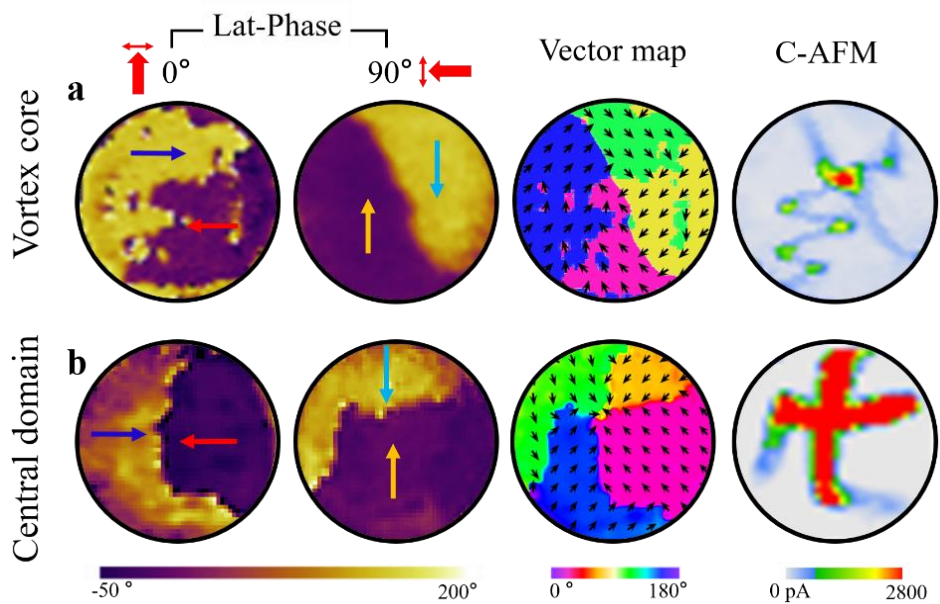

**Supplementary Figure 4 | The PFM images and domain wall conduction for the two types of topological cores,** *i.e.* vortex state (a) and center state (b), which were selected for further temperature-dependent measurements (see Fig. 2 in the main text). The thin arrows inside the images present the in-plane directions of local polarization component perpendicular to the cantilever. The diameter of each nanoisland is 400 nm.

**D. Conductivity mechanism for the cores**

To understand these emergent conduction behaviors, more measurements were performed in our experiments, including probing the temperature (T)-dependent conductivity of the core regions for the two types of nanoislands: one carries a vortex state and the other a center state (see Supplementary Fig. 4). An immediate glance of the measured current (I) - voltage (V) data at various T near room temperature suggests two distinctly different conductivity regions: the semiconducting conduction in the low bias range (V < 1.7 V) and the metallic like conduction in the high bias range (V > 1.7 V).

In proceeding, in the low bias range (V < 1.7 V), we discuss the measured data and compare the signatures of different conduction mechanisms, such as the Fowler-Nordheim mechanism [$ln(I/V^2) \sim E^{-1}$] [5], the space charge-limited model [$ln(I) \sim ln(V)$] [6], the Poole-Frenkel emission model [$ln(I/V) \sim V^{1/2}$] [7], and the Richardson Schottky Simmons emission model [$ln(I/(T^{3/2}V)) \sim V^{1/2}$] [8]. Obviously, the data can be fitted well with the Schottky model (see Supplementary Fig. 5e & 5f) for both the vortex and center cores, indicating that the Richardson-Schottky-Simmons emission is the favored mechanism accounting for the interface-limited conduction.

However, in the high bias range (V > 1.7 V), the linear I - V curves suggesting more likely the metallic behaviors can be evidenced with the positive temperature coefficient for both the vortex core and center core. Defining the bias threshold as 1.7 V, one finds the metal-like conducting behaviors above the threshold, and the thermionic behaviors [9] below the threshold, for the two types of topological cores (see Supplementary Fig. 5g & 5h).

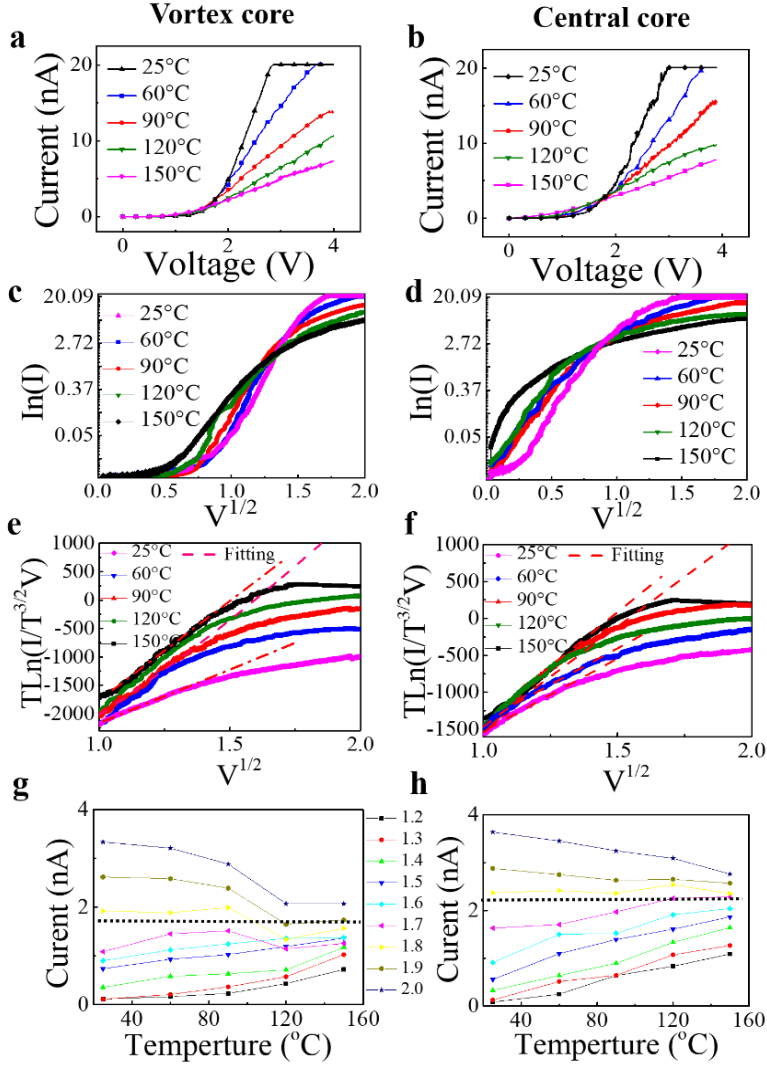

**Supplementary Figure 5 | The temperature dependent conduction behaviors for both vortex core and center core. a,b**, The I-V curves for the two types of topological cores. **c,d,** The temperature dependent I-V curves plotted as $ln$(I) ~ $V^{1/2}$ to fit the Poole-Frenkel emission model in the low bias range. **e,f,** The temperature dependent I-V curves plotted as $ln[I/(T^{3/2}V)]$ ~ $V^{1/2}$ to fit the Richardson-Schottky-Simmons emission model in the low bias range. **g,h**, The I-T curves for the two types of cores as a function of the bias voltage. The I-V curves fit well the Richardson-Schottky-Simmons emission model in the low bias range, while they conform to the metallic conducting mechanism better in the high bias range.

These observed phenomena regarding the electrical conduction of these topological cores could be explained by the band bending induced by the large net polarization-bound charge in the head-to-head domain cores (Supplementary Fig. 6). This bending leads to a large drop of conduction band below the Fermi level in the cores and thus attraction of high density of electrons responsible for the metal conductive channels, in analogue to the phenomena in charged domain walls [1, 10, 11].

When the metallic channel contacts with electrode, there usually exist a very narrow insulating gap and a wedge domain [11] in the contact region close to the electrode, giving rise to the Schottky barrier between the metallic core and electrode. This narrow gap, if not effectively overcome, would result in semiconducting behavior in the ultra-low bias range. Therefore, it is reasonable that the core region exhibits the thermionic emission behavior in the low bias range. However, in the sufficiently high bias range, the wedge domain can be annihilated and thus the bias overcomes the tunneling barrier from the insulating gap. In this case, the measured resistance comes mainly from the 1D core channel itself. As a result, the metallic conductivity becomes dominant over the interfacial thermionic effect. This is consistent with the observation of a threshold in the I-V curve, below which a nonlinear (non-Ohmic) and semiconducting behavior was identified and beyond which a liner behavior reflecting the metallic behavior was observed (see Supplementary Fig. 5a & 5b).

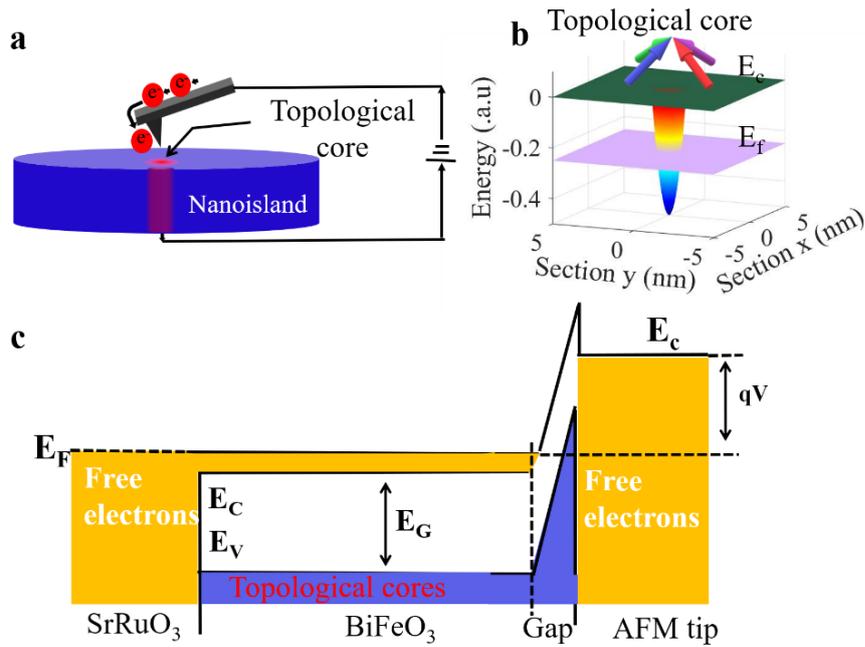

**Supplementary Figure 6 | Schematic diagrams of the band structure in the topological core region. a,b,** The band bending at the core, creating a conductive channel below the Fermi level. **c**, The band structure of the conductive core with electrode, under an electric field, wherein an insulating gap exists between the conductive channel and electrode.

### E. Evidences for unstable twisted vortex cores

A surprising observation in our experiments is the dynamic behavior of the domain structure, in particular the high sensitivity of the domain wall shape and properties near the core in response to applied electric bias. One case is the dynamic transition of the vortex core from a normal vortex state into a twisted state, as driven by an electric bias. Such a twist is basically characterized by the seriously shape-distorted domain wall from roughly straight line, while the twisted state is unstable and a recovery back to the initial state occurs gradually after the bias removal. Here, it should be mentioned that a high bias (~ 7.0 V) applied to the vortex core is sufficient for the switched distorted core state to survive for 30 min after the bias removal, until the original vortex core state is recovered, as shown in Supplementary Fig. 7. At smaller bias, the vortex core is only a transient state, which might not survive long enough for the PFM imaging. This provides dynamic evidence for the occurrence of dynamically twisted vortex core.

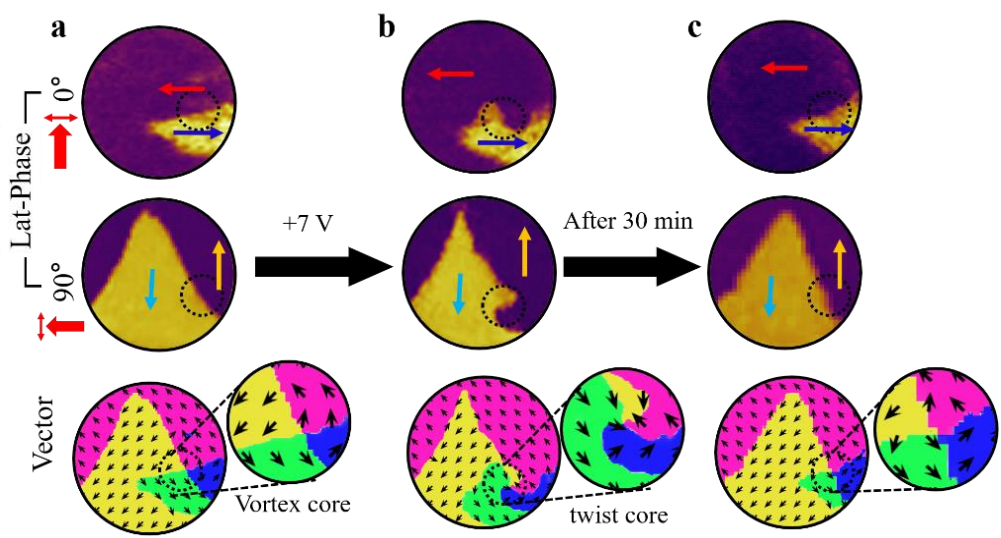

**Supplementary Figure 7 | The observation of dynamic twisted vortex core induced by applied electric field. a**, The initial vortex domain structure and corresponding CAFM map. **b**, A center-like twisted core induced by a 7.0 V bias at the vortex core, while the flux-closure domain pattern remains outside the core region. **c**, The center-like twisted core becomes unstable and returns to the initial vortex core around 30 min after the bias removal. The magnified images for the polarization vector near the vortex core and twisted core are also presented in the insets beside the vector maps. The diameter of each nanoisland is 400 nm.

**F. Lateral sizes of conductive topological cores**

An important issue is the characteristic sizes of the two types of cores, which are critical parameters for discussion. These sizes can be estamiated from the simulated condcutive mapping, noting that the simulated results agree well with our observations in terms of the transport properties and dynamic response. Nevertheless, it is impossible to give an experimental estimation directly from the PFM data since the tip used for the CAFM probing is ~ 30 nm in curvature radius which is too large for probing the dimension of those ultra-small cores.

The simulated conductivity profiles for the two types of cores are plotted in Supplementary Figure 8. By measuring the full width at half maxima (FWHM) of line profile of conductivity distribution of the topological cores extracted from the simulated data, one can derive that the core diameters for the two types of topological core are ~ 2.5 nm.

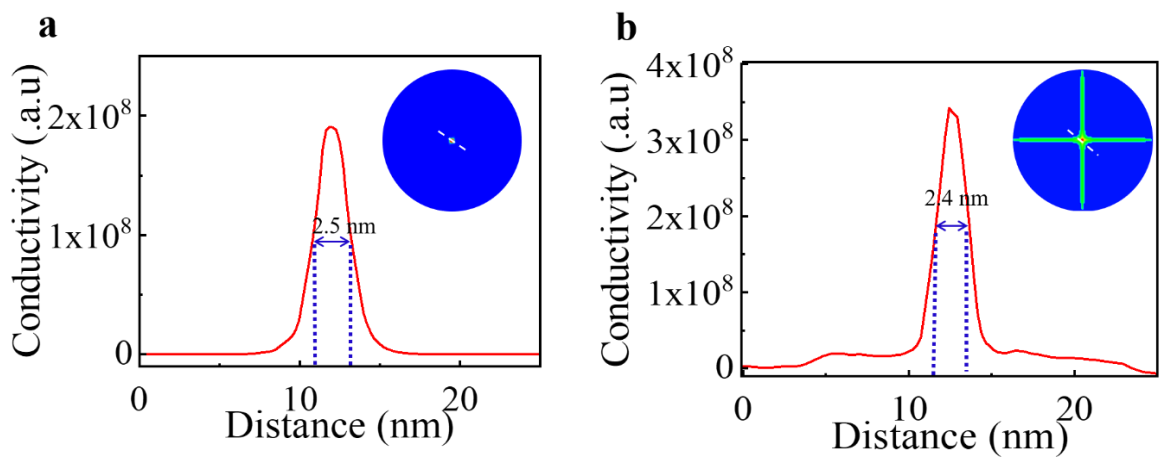

**Supplementary Figure 8 | The lateral dimension of the conductive topological cores. a,b,** The lateral dimension of the topological cores can be estimated from the full-width at half maxima of the conduction line profiles along the lines for the two topological cores, extracted

from the simulated conductivity color contours: vortex core (a) and center core (b). The diameter for each simulated map is 70 nm.

**G. Conductive core states: creation and elimination**

The detailed information for creation and elimination of the conductive core states can be found in Supplementary Fig. 9 as representative example. The initial pristine state usually exhibits a wedge-like domain pattern (with downward vertical polarization) of very low conductivity at the walls (abbreviated as wedge domain state). A vortex domain structure can be induced by applying a scanning bias of ~ 3.5V on the nanoisland, and it will convert to a center state as the bias is as high as 5.5 V and more. The center domain state can also be switched back (eliminated) to the low conductive wedge domain state by applying a negative bias of -3.5 V. This process can of course be repeatable, and again, the wedge domain state can be switched to vortex state induced by 3.5 V bias, and further applying a bias of 5.5 V leads to the formation of center state.

Selected writing and deleting of the topological states in a nanoisland array were also mapped by the PFM, as shown in Supplementary Fig. 10. It was found that initially all the nanoislands exhibit low conductive wedge domain state, and a tip scanning under a bias of 5.5 V converted all these nanoislands into the center state each. For better illustration, four nanoislands were circled as shown, and the center states were deleted and converted to the low conduction wedge domain states by the -3.5 V bias writing. Further writing by the 3.5 V bias on two selected nanoislands and 5.5 V bias on the other two nanoislands, created two vortex domain states and two center domain states, respectively. These experiments indicated that both the two types of topological states can be created and eliminated reversibly and individually.

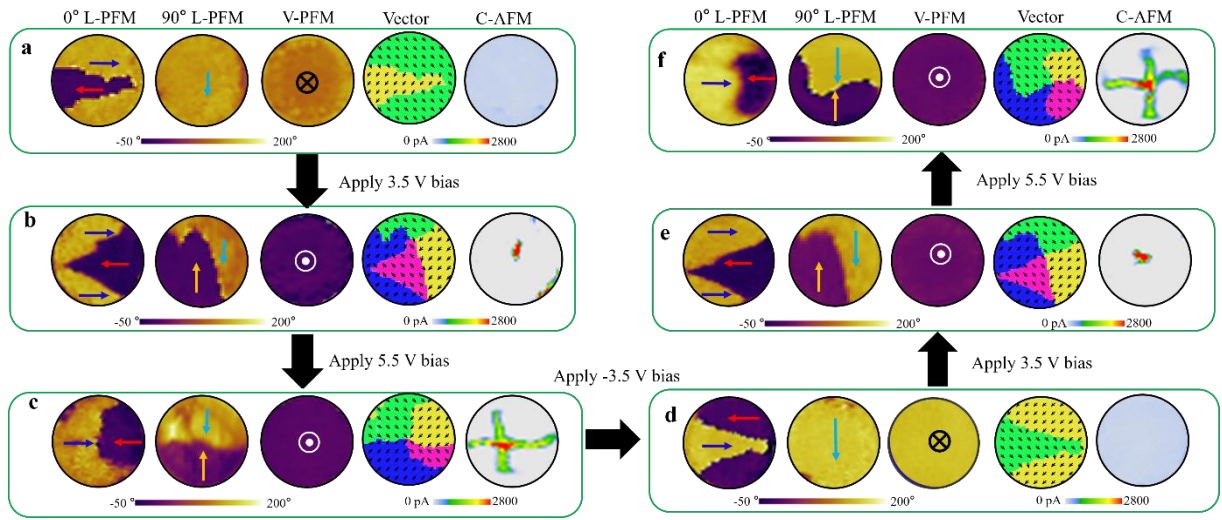

**Supplementary Figure 9 | The creation and elimination of the vortex/center states in a single nanoisland via applying scanning bias. a-f**, Here, the bias induced evolution of domain structures and corresponding conduction patterns, including: initial low conduction wedge domain state (a), creation of a vortex state by applying a bias of 3.5 V (b), creation of a center state from a vortex state by a 5.5 V bias (c), elimination of a center state (returns to the low conduction wedge domain state) by a -3.5 V bias (d), creation of a vortex state by 3.5 V (e), and conversion from vortex state to center state by 5.5 V (f). The micrographs from the left to the right are: the PFM lateral phase images (0º and 90º) used to evaluate the local polarization direction, the vertical phase images, the in-plane local polarization vector maps, and the CAFM maps. **a, b and c** share the same images as Supplementary Figure 3. The small arrows inside the PFM images present the directions of local polarization component perpendicular to the cantilever. The diameter of each nanoisland is 400 nm.

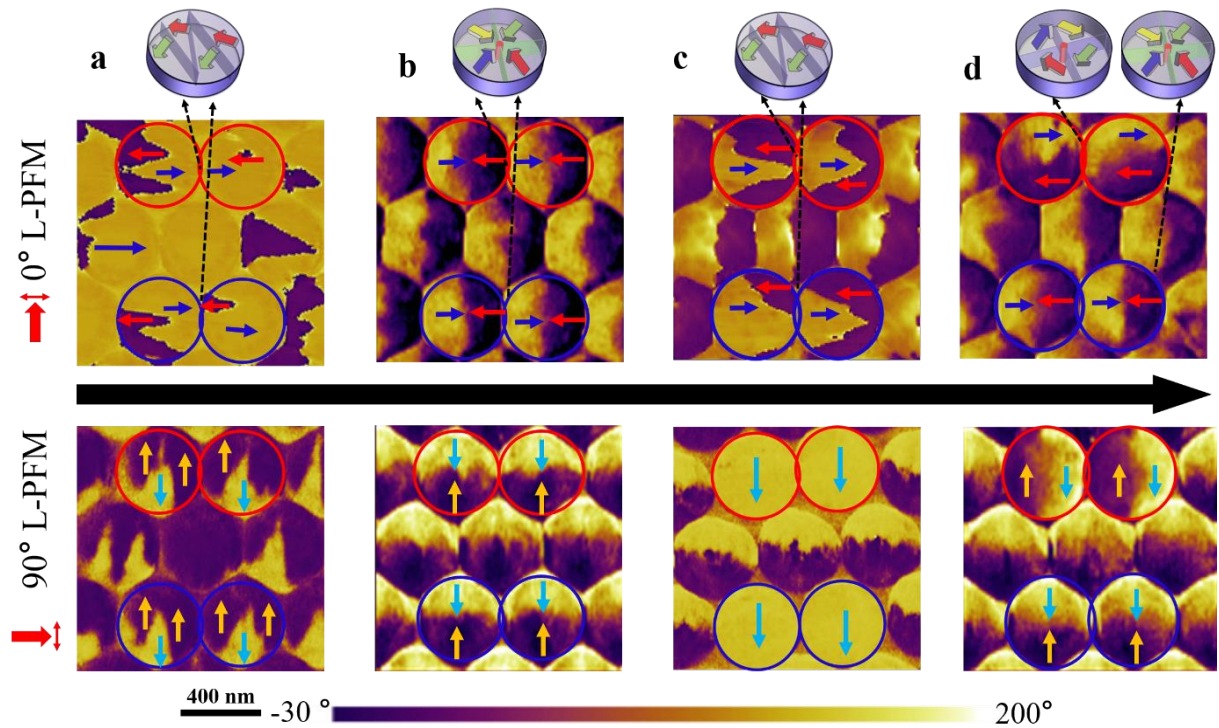

**Supplementary Figure 10 | The PFM images for creation and elimination of the topological states in a nanoisland array (corresponding to Fig. 4a in the main text). a-d**, the lateral PFM images obtained at 0º angle (top row) and 90º angle (bottom row), for various states: initial low conduction wedge domain state (a), center domain state after applying scanning bias of 5.5 V over the whole array (b), elimination of the center domain state for the four selected nanoislands back to the wedge domain state by applying -3.5 V (c), and creation of two vortex states by applying +3.5V and two center states by applying 5.5 V on the four nanoislands by re-writing (d). The four selected nanoislands are marked by the circles. The inset schematics above the PFM images present the different topological states: wedge domain, center and vortex states. The small arrows inside the PFM images present the directions of local polarization component perpendicular to the directions of cantilever.